\documentclass[draft]{agujournal2019}
\usepackage{url} 
\usepackage{amsmath,amssymb}
\usepackage[inline]{trackchanges} 
\usepackage{soul}
\drafttrue

\journalname{JGR: Space Physics}

\begin{document}

%
%

\title{On the applicability of single-spacecraft interferometry methods using electric field probes}

%
%




\authors{K. Steinvall\affil{1,2}, Yu. V. Khotyaintsev\affil{1}, D. B. Graham\affil{1}}

\affiliation{1}{Swedish Institute of Space Physics, Uppsala, Sweden}
\affiliation{2}{Space and Plasma Physics, Department of Physics and Astronomy, Uppsala University, Uppsala, Sweden}




\correspondingauthor{Konrad Steinvall}{konrad.steinvall@irfu.se}




\begin{keypoints}
\item We analyze the applicability and accuracy of three single-spacecraft interferometry methods
\item Interferometry based on probe potentials should be used only in dense plasmas and for planar waves
\item Electric-field-based interferometry in the spin-plane is accurate in all plasma conditions and for both planar and non-planar waves
\end{keypoints}

%
%

%
%


\begin{abstract}
When analyzing plasma waves, a key parameter to determine is the phase velocity. It enables us to, for example, compute wavelengths, wave potentials, and determine the energy of resonant particles. The phase velocity of a wave, observed by a single spacecraft equipped with electric field probes, can be determined using interferometry techniques. While several methods have been developed to do this, they have not been documented in detail. In this study, we use an analytical model to analyze and compare three interferometry methods applied on the probe geometry of the Magnetospheric Multiscale spacecraft. One method relies on measured probe potentials, whereas the other two use different E-field measurements: one by reconstructing the E-field between two probes and the spacecraft, the other by constructing four pairwise parallel E-field components in the spacecraft spin-plane.
We find that the potential method is sensitive both to how planar the wave is, and to spacecraft potential changes due to the wave. The E-field methods are less affected by the spacecraft potential, and while the reconstructed E-field method is applicable in some cases, the second E-field method is almost always preferable.
We conclude that the potential based interferometry method is useful when spacecraft potential effects are negligible and the signals of the different probes are very well correlated. The method using two pairs of parallel E-fields is practically always preferable to the reconstructed E-field method and produces the correct velocity in the spin-plane, but it requires knowledge of the propagation direction to provide the full velocity.
\end{abstract}

%
%

%


%
%
%
%

\section{Introduction}
Plasma waves play an essential role in the dynamics of space plasmas. Through wave-particle interactions, waves can cause, for example, plasma heating~\cite{khotyaintsev2020}, plasma transport across current layers~\cite{vaivads2004,graham2017}, anomalous resistivity~\cite{drake2003}, and mediate energy exchange between different particle species~\cite{kitamura2018}. Currently, of particular research interest is the importance of waves in the kinetic-to-global-scale chain of magnetic reconnection~\cite<>[and references therein]{khotyaintsev2019}.
In addition, observations of propagating waves can serve as diagnostics to infer instabilities and thereby plasma conditions in a distant source region. 

Electrostatic waves are found around most dynamic regions in geospace, such as the bow shock~\cite<e.g.>[]{scarf1970,balikhin2005,goodrich2018} and the reconnecting magnetopause \cite<e.g.>[]{matsumoto2003,graham2016,uchino2017}. In order to understand the cause and effect of such waves, it is important to determine their properties properties, particularly their phase velocity $\mathbf{v}$, wavevector $\mathbf{k}$, and wave potential. If the waves are observed by a multispacecraft mission such as the Magnetospheric Multiscale (MMS) mission~\cite{burch2016}, and they are larger than the inter-spacecraft separation, as is often the case when MMS is in the Earth's magnetotail, these parameters can be determined via multispacecraft interferometry~\cite{pickett2008,norgren2015,holmes2019}. However, in plasma regions such as the magnetopause and the magnetosheath where the Debye-length is short, individual electrostatic waves are rarely observed by more than a single spacecraft, necessitating the use of single-spacecraft methods to determine the wave properties. Historically, this has been accomplished using spatially separated measurements of either electron density or electric field fluctuations, as discussed in the review by~\citeA{labelle1989}. Contemporary magnetospheric and heliospheric spacecraft missions such as Cluster, MMS, and Parker Solar Probe have spatially separated electric field probes allowing only for interferometry based on measurements of electric fields or probe potentials. Several different single-spacecraft interferometry methods have been developed to analyze electrostatic waves using such measurements~\cite<e.g.>[]{vaivads2004,graham2016,vasko2018}. For example, \citeA{vasko2018} used interferometry based on probe potential signals, whereas \citeA{vaivads2004} and \citeA{graham2016} relied on interferometry using electric field measurements constructed in different ways. While these methods have been argued to be applicable in the relevant cases, their accuracy and limitations have not been documented in detail. It is the aim of the present paper to fill this literature gap using primarily numerical calculations to analyze the three different single-spacecraft interferometry methods cited above.

The outline of this paper is as follows. In Section~\ref{sec:method} we introduce and describe the three interferometry methods we will analyze. Section~\ref{sec:results} is devoted to analyzing the methods, to see how well they work under different circumstances. In Section~\ref{sec:spacecraft} we discuss the implications our results have on spacecraft analysis, and present two examples of in-situ observations of solitary waves on which we apply two of the methods. Finally, in Section~\ref{sec:summary} we summarize our results and conclusions.

\section{Methods}
\label{sec:method}
\subsection{Data Description}
\label{sec:data_theory}
\begin{figure}
\noindent\includegraphics[width=0.5\textwidth]{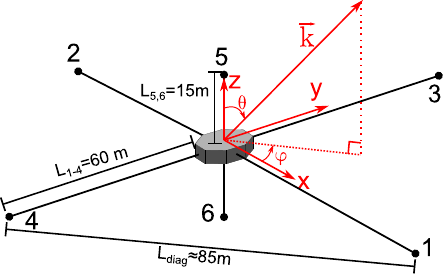}
\caption{Schematic showing the probe geometry of MMS. Not to scale.}
\label{fig:probes}
\end{figure}

We investigate three interferometry methods using an analytical electrostatic model that we solve numerically. Our approach is straight-forward: We define an electrostatic potential profile, give it a phase velocity, and let it propagate past seven measurement points, corresponding to the six electric field probes and the spacecraft body. Using synthetic time series data from these points, we apply the different interferometry methods and compare their resulting velocity to the prescribed phase velocity. We consider the MMS spacecraft~\cite{burch2016}, which has four electric field probes in the spin plane~\cite{lindqvist2016}, each at a distance of 60 meters from the spacecraft, and two electric field probes along the spin-axis~\cite{ergun2016} each located approximately 15 meters from the spacecraft (the tip-to-tip distance is 30.4 meters). We label the probes according to the schematic in Fig.~\ref{fig:probes}, so that probes 1-4 are in the spin-plane, and 5-6 are along the spin-axis. We define a coordinate system where $\hat{x}$ is along probes 1-2, $\hat{y}$ is along 3-4, and $\hat{z}$ is along 5-6, as illustrated. In addition, we define the polar angle $\theta$ to be measured from the $z$-axis, and the azimuthal angle in the $xy$-plane, $\varphi$, to be measured from the $x$-axis. This is the coordinate system that we use throughout this work. The actual MMS spacecraft is spinning with a period of 20 seconds, but as we are interested in wave phenomena on millisecond scales, we neglect spin-related effects. 

It is important that our model treats the same type of data that is obtained by spacecraft in space. For this reason we devote a couple of paragraphs to describing how the measured potentials and electric fields are related to the physical electrostatic potential and electric field in the plasma.  For convenience, we summarize the important quantities in Table~\ref{tab:quantities}.
 \begin{table}
 \caption{Quantities used in our calculations}
 \centering
 \label{tab:quantities}
 \begin{tabular}{llcc}
 \hline
  Symbol  & Name & Measured/Not measured & Equation \\
 \hline
    $\Phi(\mathbf{r},t)$  & Wave electrostatic potential & Not measured & -\\
   $\Phi(\mathbf{r}_{i/\text{sc}})$  & Wave electrostatic potential at probe $i$/the spacecraft & Not measured & -\\
   $\Phi_{p,i}$  & Potential of probe $i$ & Not measured & -\\
   $\Phi_\text{sc}$  & Spacecraft potential & Not measured & -\\   
   $\delta\Phi_\text{sc}$  & Spacecraft potential fluctuation & Not measured & Eq.~(\ref{eq:scpot_fluct})\\
   $V_i$  & Voltage measurement of probe $i$ & Measured & Eq.~(\ref{eq:V})\\
   $E_{ij}$  & Electric field from probes $i$ and $j$ & Measured & Eq.~(\ref{eq:E})\\
 \hline
\end{tabular}
\end{table}

We start with potential measurements. What electric field probes actually measure is the potential difference between the probe potential $\Phi_{p,i}$ and the spacecraft potential $\Phi_\text{sc}$, that is, 
\begin{equation}
\label{eq:voltage_def}
    V_i=\Phi_{p,i}-\Phi_\text{sc}.
\end{equation}
We refer to these measured potential differences as \emph{voltages}, to distinguish them from the various potentials which cannot be measured directly. However, the physically interesting quantities are not $\Phi_{p,i}$ and $\Phi_\text{sc}$ (and the corresponding $V_i$), but rather the difference between the electrostatic potential of the wave at the locations of the probes and the spacecraft, $\Phi(\mathbf{r}_i)$ and $\Phi(\mathbf{r}_\text{sc})$. These latter quantities (and their difference) cannot be measured directly using electric field probes.

So, how do potentials of the probes and spacecraft, $\Phi_{p,i}$ and $\Phi_\text{sc}$, relate to $\Phi(\mathbf{r}_i)$ and $\Phi(\mathbf{r}_\text{sc})$? To answer this question it can be instructive to understand why these potentials differ. There are two primary reasons for why, in general, $\Phi_{p,i}\neq\Phi(\mathbf{r}_i)$. The first reason is that although the probe is a good conductor, it is not perfectly grounded to the plasma. This is because its potential is changed by the continuous absorption and emission of charges until the total currents to and from the probe are balanced. The result of this process is a potential difference between the probe and plasma, called the probe-to-plasma potential. The probe-to-plasma potential can be controlled (reduced) by applying a bias-current to the probe, so that $\Phi_{p,i}$ is close to the plasma potential (typically, it can differ by about $\sim1$ V). More importantly, biasing the probe makes the probe-to-plasma potential insensitive to current fluctuations~\cite{fahleson1967,mozer2016}, which could, for example, be due to the density fluctuations $\delta n_e$ associated with electrostatic waves, $\delta n_e=\epsilon_0\nabla^2\Phi/e$. Since we are presently interested in the analysis of waves, such a constant probe-to-plasma potential does not affect the analysis, and we do not need to consider it in our model. Thus, we consider $\Phi_{p,i}$ to be given by the plasma potential at the location of the probe. This point brings us to the second reason why $\Phi_{p,i}\neq\Phi(\mathbf{r}_i)$, namely that the presence of the spacecraft modifies the potential structure in its vicinity, thereby contributing to $\Phi_{p,i}$.

Similar to the probes, the potential of the spacecraft body is determined by the currents. In contrast to the probes however, we cannot bias the spacecraft body. This has the consequence that $\Phi_\text{sc}$ is affected by $\delta n_e$. A simplified expression for the corresponding potential fluctuation $\delta\Phi_\text{sc}$ can be derived from the current balance condition, $I_\gamma+I_e=0$, where $I_\gamma$ and $I_e$, are the photo- and plasma electron currents to the spacecraft respectively. Implicit in this condition is the assumption that the spacecraft is sunlit, and that any additional currents, e.g. due to plasma ions and secondary electrons are negligible. We are thus considering the case when active spacecraft potential control (ASPOC)~\cite<e.g.>[]{torkar2016} is turned off. For the case of a Maxwellian plasma, $I_\gamma$ and $I_e$ are, respectively, given by
\begin{eqnarray}
\label{eq:currents}
    I_\gamma = I_{\gamma 0}\exp\left(\frac{-e\Phi_\text{sc}}{T_\gamma}\right), & I_e = -en_eS\left(\frac{T_e}{2\pi m_e}\right)^{1/2}\left(1+\frac{e\Phi_\text{sc}}{T_e}\right),
\end{eqnarray}
where $I_{\gamma 0}$ is an amplitude coefficient, $n_e$ and $m_e$ are the electron density and mass respectively, $S$ is the spacecraft surface area, and $T_\gamma$ and $T_e$ are the photo- and plasma electron temperatures, respectively \cite<e.g.>[]{mott_smith1926,pedersen1995,roberts2020}.
Setting $I_\gamma+I_e=0$ and assuming $T_e\gg e\Phi_\text{sc}$, we find
\begin{equation}
\label{eq:scpot_fluct}
    \delta\Phi_\text{sc} = -\frac{T_\gamma}{e} \ln{\left(1+\frac{\delta n_e}{n_e}\right)}.
\end{equation}
So while the DC spacecraft potential offset with respect to the plasma can be ignored in the context of wave analysis, $\delta\Phi_\text{sc}$ can be very important. The spacecraft potential thus results in an electrostatic potential field which decreases with an increasing distance from the spacecraft. The booms to which the probes are attached are grounded to the spacecraft, and thus held at the spacecraft potential~\cite{cully2007}. So even though the probes are located relatively far from the spacecraft body (up to 60 meters from the spacecraft), the booms bring the spacecraft potential close to the probes, and the probes sense a fraction $0\leq\xi\leq1$ of $\Phi_\text{sc}$. This means that even if no electrostatic wave is present, perturbations in the plasma density will cause a voltage $\delta V_i=\Phi_{p,i}-\Phi_\text{sc}=-(1-\xi_i)\delta\Phi_\text{sc}$.
When an electrostatic wave with an associated potential difference $\Phi(\mathbf{r}_{i})-\Phi(\mathbf{r}_\text{sc})$ between the location of the probe and spacecraft is present, it follows from superposition that the measured $V_i$ relates to the electrostatic potential in the plasma as
\begin{equation}
\label{eq:V}
    V_i=\Phi(\mathbf{r}_i)-\Phi(\mathbf{r}_\text{sc})-(1-\xi_i)\delta\Phi_{\text{sc}}
\end{equation}

Experimentally determining the $\xi$-parameters is complicated, but \citeA{graham2018b} did so for MMS using the spin-plane average probe-to-spacecraft potential $V_{\text{psp}}=(V_1+V_2+V_3+V_4)/4$ and electron data. They reported a relation $\Phi_\text{sc}=-1.2V_\text{psp}+V_c$, where $V_c$ is a spacecraft specific constant and the factor 1.2 corresponds to $(\xi_1+\xi_2+\xi_3+\xi_4)/4=0.167$. Due to the symmetry of the MMS spacecraft and their electric field probes, it is reasonable to assume $\xi_1=\xi_2=\xi_3=\xi_4\equiv\xi_{1-4}$ and $\xi_5=\xi_6\equiv\xi_{5,6}$.  No similar analysis for the axial probes has been published, and $\xi_{5,6}$ is unknown. However, since the axial probes are closer to the spacecraft, we expect $\xi_{5,6}\geq\xi_{1-4}$. Naturally, $\xi$ depends on the local Debye length, and the value of \citeA{graham2018b} was derived in the magnetosphere, where the Debye length is typically larger than the the distance between the probes. In a plasma with shorter (longer) Debye length, we expect $\xi$ to be smaller (larger). Because of this variability, we will treat $\xi_i$ as a free parameter in our calculations. 

Another way in which the spacecraft affects the surrounding potential is via the emission of photoelectrons. Because the spacecraft is only partially illuminated the emitted photoelectrons will form a cloud that is asymmetric with respect to the spacecraft body~\cite{mozer1978,cully2007}, giving rise to a sunward electric field. If this cloud extends far enough from the spacecraft, it will affect the probe potentials, giving the voltages a spin-phase dependence. However, since we are interested in wave analysis on millisecond scales, we can consider this to be another DC offset which does not affect our analysis. 

The plasma environment can also be disturbed by the formation of wakes in the vicinity of the spacecraft, which give rise to an electric field~\cite{andre2021}. If a wake is wide, the probe signals will vary slowly on a scale of seconds, and we can treat this effect as a DC offset, so it will not affect the interferometry results. On the other hand, narrow wakes of large amplitude can be more problematic, since sharper wake boundaries imply shorter time-scale changes that can interfere with the analysis. Such wakes can occur in the solar wind, and it is therefore important to ensure that the probes are not located near the expected wake when applying interferometry. However, since the wakes are narrow, the probes only spend a very small fraction of their time in the wake, enabling interferometry during most of the spin.

For the MMS spacecraft, there is another effect similar to narrow wakes which can influence the measurements, namely partial shadowing of a spin-plane probe by the axial booms. Such shadowing causes a significant reduction in photoemission, affecting the probe-plasma coupling and the probe potential. Again, these shadows are narrow, which means that they can interfere with interferometry. Shadow occurrences are well known and indicated in the Electric field Double Probe (EDP) data products, which makes it easy to avoid them.

Electric field measurements are comparatively more straight-forward to understand. They can be obtained by taking the difference of two probe voltages and dividing by the distance between the probes $L_{ij}=|\mathbf{r}_i-\mathbf{r}_j|$, i.e.
\begin{equation}
\label{eq:E}
E_{ij}=-\frac{V_j-V_i}{L_{ij}}=-\frac{\Phi(\mathbf{r}_j)-\Phi(\mathbf{r}_i)-\Delta\xi\delta\Phi_\text{sc}}{L_{ij}},
\end{equation}
where $\Delta\xi=\xi_i-\xi_j$, and we have used Eq.~(\ref{eq:V}) for the second equality. If $\xi_j=\xi_i$, i.e. if the probes are identical and placed symmetrically with respect to the spacecraft potential, $E_{ij}$ reduces to $E_{ij}=\left[\Phi(\mathbf{r}_i)-\Phi(\mathbf{r}_j)\right]/L_{ij}$. In this case, the shape of $E_{ij}$, which is determined by $\Phi(\mathbf{r}_i)-\Phi(\mathbf{r}_j)$ is accurately resolved (in the large wavelength limit). However, the amplitude of this field can still deviate from the physical field due to the conducting spacecraft and booms. They cause the plasma electric field to locally short circuit, leading to a reduction in the effective length of the antenna, $L_\text{eff}$, \cite<See Fig. 5 in>[]{pedersen1998}. This means that $L_{ij}$ is not the correct length-scale of the E-field, and the measured E-field magnitude is affected. The degree to which $L_\text{eff}$ varies is different for different spacecraft and plasma conditions~\cite{khotyaintsev2014,mozer2020,steinvall2021b}. However, since only the amplitude of $E_{ij}$ is affected, the time delay obtained from interferometry remains unchanged.

\subsection{Model Setup}
Now that the we have defined the relation between measured and physical quantities, we describe the analytical model which we use to analyze the interferometry methods.
We define seven points in 3D-space corresponding to the position of the six probes and the spacecraft as in Fig.~\ref{fig:probes}, with the spacecraft positioned at the coordinate origin. Since the probes and the spacecraft are small compared to the 15 and 60 meter booms, we simplify the problem by treating the probes and spacecraft as points (no volume), and let their potential respond instantaneously to changes in the ambient electrostatic potential at their location. By prescribing and evaluating some arbitrary potential profile $\Phi(\mathbf{r},t)$, at the location of the probes and the spacecraft, we can then construct the measured voltages and electric fields using Eqs.~(\ref{eq:V}) and~(\ref{eq:E}). We give the potential structure some phase velocity $\mathbf{v}'$ and let it propagate over our measurement points (our "spacecraft"), giving us a set of synthetic time-series data. By applying the different interferometry methods to the synthetic data, we investigate how well the methods reproduce the prescribed velocity under different circumstances. In particular, we investigate how the angle $\alpha$ between the estimated velocity, $\mathbf{v}$, and $\mathbf{v}'$, and the speed ratio $v/v'$ depend on the direction of the wavevector, the three-dimensional profile of the wave, and spacecraft potential effects. To focus on the methods themselves, we use a very high temporal resolution (or, equivalently, very slow phase velocities). This means that we do not consider the errors introduced due to the finite sampling rate of the instrument, since such errors affect all methods in the same way. For the same reason, we do not investigate any effects due to short wavelengths. We focus on the case of solitary waves with Gaussian potential profiles since this allows us to include the effects of non-planar structures. However, the case of sinusoidal plane waves is also discussed in the text.

In the following, we present three single-spacecraft interferometry methods that have been used for wave analysis in space plasmas. One is based on the probe voltages, and the remaining two use electric field measurements in two different ways.

\subsection{Voltage Interferometry}
\begin{figure}
\noindent\includegraphics[width=\textwidth]{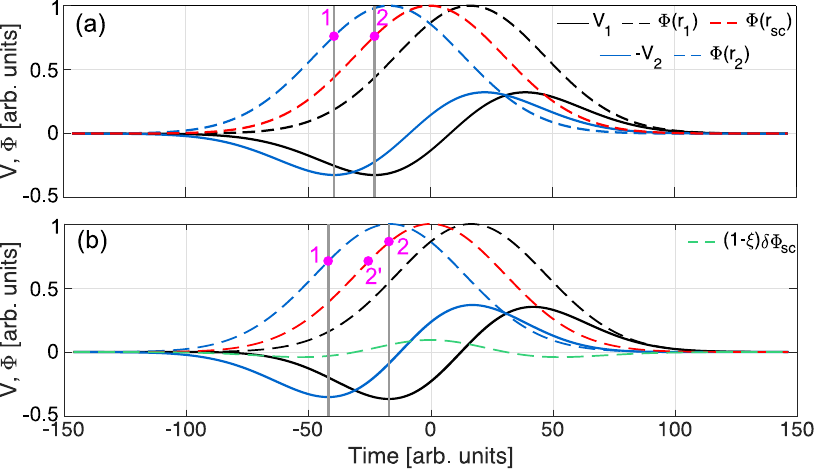}
\caption{Voltage interferometry examples. (a) Neglecting spacecraft potential effects. (b) Including spacecraft potential effects. The vertical grey lines indicate the first peak of the $V$ signals. The magenta points marked 1 and 2 correspond to the value of $\Phi(\mathbf{r}_2)$ and $\Phi(\mathbf{r}_\text{sc})$ during those peaks. The point 2' in (b) corresponds to the same phase as point 1, to highlight the observed voltage phase-shift.}
\label{fig:Vmethod}
\end{figure}

Interferometry based on the measured voltages (or monopolar E-field measurements) has been used to analyse waves in the Earth's high-altitude polar magnetosphere~\cite{franz1998}, at the Earth's bow shock \cite{vasko2018,vasko2020,wang2021}, and in the solar wind \cite{mozer2021}. Most of these studies have applied the analysis to observations of solitary waves, but \citeA{mozer2021} also used it to analyze more sinusoidal waves. We will refer to this as the VI (voltage interferometry) method. To illustrate the VI method, we apply our analytical model to a solitary wave characterized by a cylindrically symmetric double Gaussian potential
\begin{equation}
\label{eq:EH}
    \Phi(x',y',z',t)=\Phi_0 \exp\left(-\frac{x'^2+y'^2}{2L_\perp^2}\right)\exp\left(-\frac{(z'-v't)^2}{2L_\parallel^2}\right),
\end{equation}
where the primed coordinate system has its origin at the center of the Gaussian potential, $L_\perp$ and $L_\parallel$ are the length scales perpendicular and parallel to the direction of propagation, $\hat{z}'=\hat{k}$. In Fig.~\ref{fig:Vmethod}a we present synthetic data of the relevant voltages and potentials for probes 1 and 2. For the data we use $L_\parallel=50$ m, $L_\perp=50L_\parallel$, and $\hat{k}$ is specified by $\theta=40^\circ$, $\varphi=45^\circ$. For simplicity, we temporarily ignore any effects due to $\delta\Phi_\text{sc}$ and set $\xi_{1-4}=\xi_{5,6}=0$. 
The ultimate goal is to determine the time delay between the potentials (dashed lines), as they correspond to the motion of the actual structure, but to do this, we rely on the measured voltages $V_1$ and $-V_2$ (solid lines). Upon inspection, we find that the first peaks in $V_1$ and $-V_2$ (marked by the vertical grey lines) occur during the same phase of $\Phi_\text{sc}$ and $\Phi(\mathbf{r}_2)$ respectively, as highlighted by the magenta dots marked 1 and 2. This means that the time delay between $V_1$ and $-V_2$, $\Delta t_{12}$, is the same as that between $\Phi(\mathbf{r}_2)$ and $\Phi(\mathbf{r}_\text{sc})$ (and also between $\Phi(\mathbf{r}_\text{sc})$ and $\Phi(\mathbf{r}_1)$).
By repeating the analysis for the other probe pairs, we obtain $\Delta t_{34}$ and $\Delta t_{56}$. The velocity of the structure can then be computed as~\cite{vasko2018}
\begin{eqnarray}
\label{eq:V_vel}
    v=\left(\Delta t_{12}^2/L_1^2+\Delta t_{34}^2/L_3^2+\Delta t_{56}^2/L_5^2\right)^{-1/2}, & \hat{k}_{ji}=v\Delta t_{ij}/L_{i},
\end{eqnarray}
where $\hat{k}_{ji}$ is the component of the wave unit vector in the direction of probes $ji$, i.e. $\hat{k}_{21}\parallel\hat{x}$, $\hat{k}_{43}\parallel\hat{y}$, and $\hat{k}_{65}\parallel\hat{z}$.

To illustrate the effect of the complications related to $\Phi_\text{sc}$, we repeat the analysis for the same electrostatic potential, but with $\xi_{1-4}=0.167$, and with $\delta\Phi$ given by Eq.~(\ref{eq:scpot_fluct}) using $\Phi_0=1$ V, $T_\gamma=5$ eV, and $n_e=1$ cm$^{-3}$. The resulting $V$ and $\Phi$ signals are shown in Fig.~\ref{fig:Vmethod}b. In this case the peaks of $V_1$ and $-V_2$ no longer correspond to the same phase of $\Phi_\text{sc}$ and $\Phi(\mathbf{r}_2)$. This is illustrated by the magenta dots marked $1$ and $2'$, which show respectively the phase of $\Phi(\mathbf{r}_2)$ at the peak of $-V_2$, and the corresponding phase of $\Phi_\text{sc}$. The time delay between $1$ and $2'$, which is the time delay related to the actual velocity of the wave, is approximately half as long as the one estimated from the $V$ measurements (i.e. the delay between dots $1$ and $2$).
The source of this phase shift is the $(1-\xi)\delta\Phi_\text{sc}$ contribution to $\Phi_{p,i}$, plotted in green. While $\Phi(\mathbf{r}_\text{sc})$ is symmetric with respect to $\Phi(\mathbf{r}_1)$ and $\Phi(\mathbf{r}_2)$, $\Phi(\mathbf{r}_\text{sc})+(1-\xi)\delta\Phi_\text{sc}$ (which is what $\Phi(\mathbf{r}_i)$ is subtracted by to form $V_i$) is not. This broken symmetry leads to the observed voltage phase shifts. Similar phase shifts occur when the potential structure becomes more three-dimensional, as is later shown in Section~\ref{sec:voltage_stats}. 

Problems due to the changes in the spacecraft potential such as the one discussed above are inherent to the methods relying on these voltage measurements, as they are measured with respect to the variable spacecraft potential. We can avoid such problems by instead relying on electric field interferometry.

\subsection{Electric Field Interferometry}
Since electric fields are calculated from the difference between two voltage measurements, any spacecraft potential contribution in the voltages is canceled if the two probes have the same $\xi$ as shown in Eq.~(\ref{eq:E}). Because of this, electric field based interferometry has been used to analyze waves in various contexts \cite<e.g.>[]{vaivads2004,balikhin2005,khotyaintsev2010,graham2016,steinvall2021}. We focus on two interferometry methods using electric field measurements.

\subsubsection{Reconstructed Electric Fields}
\label{sec:method_E60}
\begin{figure}
\noindent\includegraphics[width=\textwidth]{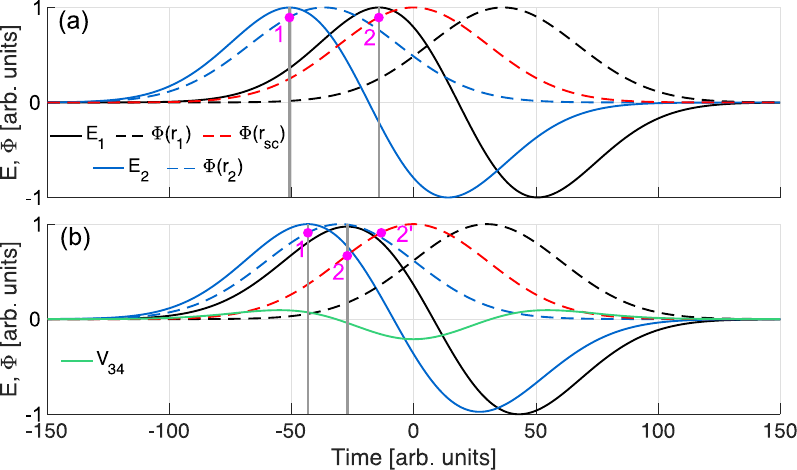}
\caption{Reconstructed electric field interferometry examples for two values of $\gamma$. (a) $\gamma=0^\circ$ ($\theta=90^\circ$, $\varphi=0^\circ$). (b) $\gamma=35^\circ$ ($\theta=90^\circ,\varphi=35^\circ$). Same format as Fig.~\ref{fig:Vmethod}, but with $E$ instead of $V$.}
\label{fig:Emethod}
\end{figure}
The first E-field method, used by for example~\citeA{khotyaintsev2010} and \citeA{graham2016}, attempts to avoid these spacecraft potential problems by reconstructing the electric field between the individual probes of a probe pair and the spacecraft, using voltage averaging. We will refer to this method as the reconstructed E-field interferometry (REI) method. The principle behind this method is as follows. When $\mathbf{v}$ (and thus also $\mathbf{k}$) is close to aligned with one of the probe pairs, e.g. probes $1$ and $2$, it is close to orthogonal to probes $3$ and $4$. Since $\mathbf{k}\times\mathbf{E}=0$ for electrostatic plane waves, the wave potential $\Phi$ does not vary much between probes 3 and 4 in the above scenario, and the voltage average $(V_3+V_4)/2$ is  a good approximation of a hypothetical voltage measurement made at $\mathbf{r}_\text{sc}$, $V_\text{sc}$. Individual electric fields between each probe and the spacecraft can then be calculated as $E_1=(V_\text{sc}-V_1)/L_1$ and $E_2=(V_2-V_\text{sc})/L_2$. These E-fields are aligned with, and spatially separated along the direction defined by the probe pair. Interferometry can then be applied to $E_1$ and $E_2$ to find the time delay and ultimately the wave speed along the probe pair. Using Eq.~(\ref{eq:V}) one can show that $E_1$ and $E_2$ are unaffected by spacecraft potential effects when $\xi_1=\xi_2=\xi_3=\xi_4$. The method is expected to be most accurate when $\mathbf{k}$ is parallel to a probe pair, since this yields the longest baseline for interferometry and the most reliable estimate of $V_\text{sc}$.

To illustrate this method, we take the same solitary wave as before, with $\mathbf{v}$ being at a small angle $\gamma$ to probes 1 and 2. We compute $V_\text{sc}$ and construct $E_1$ and $E_2$. In Fig.~\ref{fig:Emethod} we plot $E_1$ and $E_2$ for $\gamma=0^\circ$ in \ref{fig:Emethod}a, and $\gamma=35^\circ$ in \ref{fig:Emethod}b. Similar to the VI results in Fig.~\ref{fig:Vmethod}a, the peaks in $E_2$ and $E_1$ occur during the same phase of $\Phi(\mathbf{r}_2)$ and $\Phi(\mathbf{r}_\text{sc})$, meaning that the observed $\Delta t_{12}$ between the $E$ signals is the same as that between the potentials, i.e. $\Delta t_{12}$ is indeed accurate. The velocity component along the probe pair is then given by $v_{12}=L_1/\Delta t_{12}$. If $\hat{k}$ (and subsequently $\gamma$) is a priori known, e.g. from the assumption of magnetic-field-aligned propagation or maximum variance analysis for the case of plane waves, we can estimate the total speed as~\cite{graham2016}
\begin{equation}
\label{eq:v_E60}
    v = v_{12}\cos\gamma.
\end{equation}
However, as we go to larger $\gamma$ in Fig.~\ref{fig:Emethod}b, we again find a significant phase shift in $E_1$ and $E_2$, which leads to an erroneous $\Delta t_{12}$. This phase shift is due to $V_{34}$ no longer being an accurate estimate of the voltage at the spacecraft position.

It is important to note that this method requires that $\hat{k}$ is known, as this determines the choice of probes used for interferometry and for calculating $V_\text{sc}$. The appropriate choice is to use the probe pair most aligned with $\hat{k}$ for the interferometry~\cite{graham2016}. Moreover, as illustrated in Fig.~\ref{fig:Emethod}b, it is important that $\hat{k}$ is close to perpendicular to a probe pair, as only then will their average voltage give a good representation of $V_\text{sc}$.

One important problem with this method arises when $\mathbf{k}$ is assumed to be close to aligned with the axial probes. We then use $V_\text{sc}=V_{1234}=(V_1+V_2+V_3+V_4)/4$. Due to the different boom lengths of the axial and spin-plane probes, $\xi_{1-4}$ is in general expected to be smaller than $\xi_{5,6}$. When computing $E_5$ and $E_6$, this $\xi$ difference introduces additional spacecraft potential contributions $\mp(\xi_{5,6}-\xi_{1-4})\delta\Phi_\text{sc}/L_{5,6}$, to $E_{5}$ ($-$) and $E_6$ ($+$), respectively (see Eq.~(\ref{eq:E})).

\subsubsection{Diagonal Electric Fields}
\label{sec:DEI_intro}
\begin{figure}
\noindent\includegraphics[width=\textwidth]{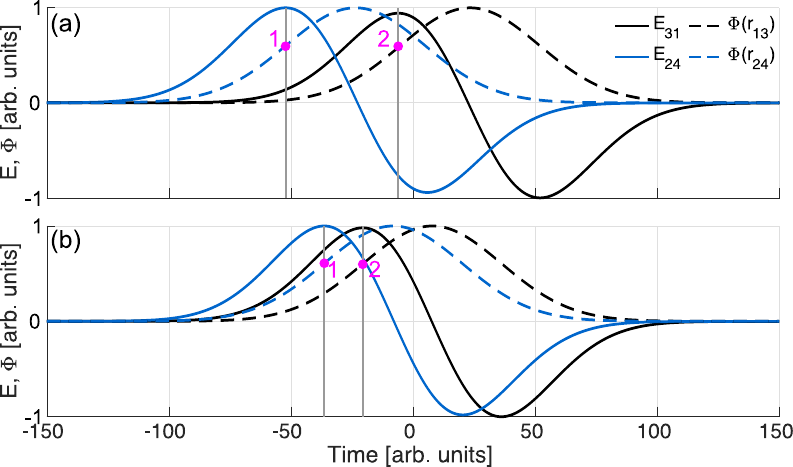}
\caption{Diagonal electric field interferometry examples for two different propagation directions. (a) $\varphi=20^\circ$, $\theta=90^\circ$. (b) $\varphi=20^\circ$, $\theta=20^\circ$. Same format as Fig.~\ref{fig:Emethod}.}
\label{fig:Ediag}
\end{figure}

The second E-field interferometry method, which has been used by e.g.~\citeA{vaivads2004} and \citeA{balikhin2005}, is limited to the spin-plane. It avoids the issue of a variable spacecraft potential by using electric fields between probes rotated 90$^\circ$ about the spacecraft from each other, instead of the conventional 180$^\circ$. We refer to this method as the diagonal E-field interferometry (DEI) method. This way of constructing the E-fields provides two sets of two electric field components pointing in the same direction. These components are $E_{41}\parallel E_{23}$, and $E_{13}\parallel E_{42}$, following the notation of Eq.~(\ref{eq:E}) (see Fig.~\ref{fig:probes} for probe positions). Their baseline and spatial separation are both $L_{\text{diag}}=\sqrt{2}L_{1-4}\approx85$m. The respective time delays between the field components are $\Delta t_{41,23}$ and $\Delta t_{42,13}$. Taking $\Delta t_{41,23}$ as an example, it corresponds to the velocity component along the displacement vector between the $E_{41}$ and $E_{23}$ measurements, which is orthogonal to the E-components themselves. This has the implication that the method does not work when $\hat{k}$ is perfectly along one of the diagonal E-field component pairs since, in that case, there is no propagation between the measurement points. However, as long as $\hat{k}$ is at an angle to both pairs, $\Delta t_{41,23}$ and $\Delta t_{42,13}$ can be determined. Once obtained, these time delays can be used to calculate the phase velocity in the spin-plane as
\begin{eqnarray}
\label{eq:v_E85}
  v = (\Delta t_{41,23}^2/L_{\text{diag}}^2 +\Delta t_{42,13}^2/L_{\text{diag}}^2)^{-1/2}, & \hat{k}=
  \begin{pmatrix}
  \frac{1}{\sqrt{2}} & -\frac{1}{\sqrt{2}} \\
  \frac{1}{\sqrt{2}} & \frac{1}{\sqrt{2}}
  \end{pmatrix}
  \begin{pmatrix}
  \Delta t_{42,13} \\
  \Delta t_{41,23} \\
  \end{pmatrix}
  v/L_\text{diag},
\end{eqnarray}
where we use the rotation matrix to return to the original coordinate system defined in Fig.~\ref{fig:probes}. In Fig.~\ref{fig:Ediag}, we show $E_{13}$ and $E_{42}$ as well as $\Phi$ evaluated at the midpoint between the probes for two propagation directions $\theta=90^\circ, \varphi=20^\circ$ (a), and $\theta=20^\circ, \varphi=20^\circ$ (b). For both cases the time delay between $E_{13}$ and $E_{42}$ is the same as that between $\Phi(\mathbf{r}_{13})$ and $\Phi(\mathbf{r}_{24})$, resulting in an accurate velocity estimate in the spin-plane.
If $\hat{k}$ is known, and is specified by some polar angle $\theta$, we can compute the total speed by multiplying the spin-plane velocity by $\sin\theta$.

\section{Statistical Results}
\label{sec:results}
Next, we investigate the accuracy of the three methods statistically. We focus on solitary waves, but the main conclusions are also true for plane waves (with sufficiently long wavelengths to avoid spatial aliasing) as will be discussed. The trajectory of the solitary wave is set so that its center passes the point corresponding to the spacecraft body.

\subsection{Voltage Interferometry}
\label{sec:voltage_stats}
Since the VI method is affected by spacecraft potential effects, we divide the analysis into two parts. In the first part, we set $\delta\Phi_\text{sc}=0$, corresponding to the case of a negligible $\delta n_e\ll n$. In the second part, we include such spacecraft potential effects and investigate how the results are affected.

\begin{figure}
\noindent\includegraphics[width=\textwidth]{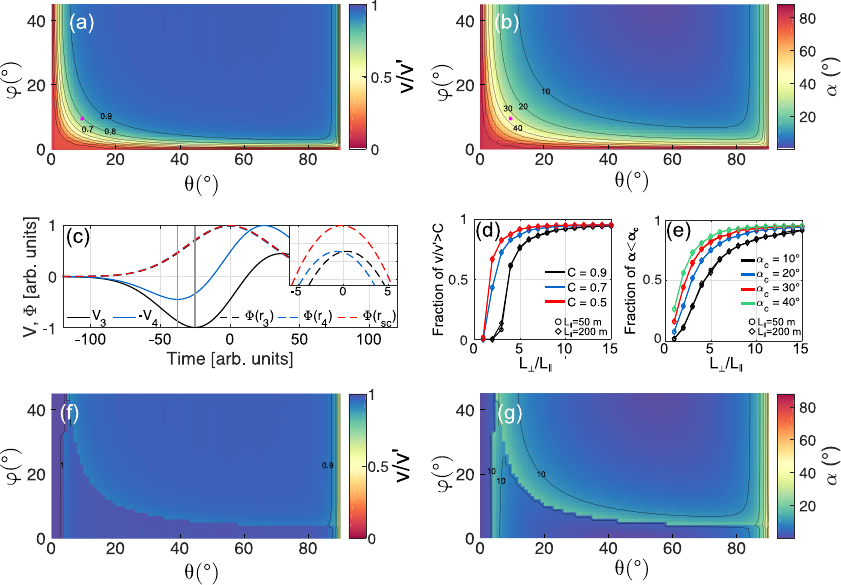}
\caption{Numerical results for voltage interferometry ignoring spacecraft potential effects. (a-b) $v/v'$ and $\alpha$ as a function of $\theta$ and $\varphi$ for a fixed $L_\perp/L_\parallel=5$. (c) Example of time series data showing a large voltage phase shift ($\theta=10^\circ, \varphi=10^\circ$). (d-e) The fraction of points on the surfaces in (a) and (b) fulfilling $v/v'>C$ and $\alpha<\alpha_c$, respectively, as a function of $L_\parallel$ and $L_\perp/L_\parallel$ for three values of $C$ and four values of $\alpha_c$. Note that the diamonds and circles are almost perfectly overlapping. (f-g) Same format as (a-b), except we set $\Delta t_{ij}=0$ if the voltages are significantly distorted (see text).}
\label{fig:V_L_stats}
\end{figure}

Due to the fact that there is only a finite number of measurement points, it is not obvious whether or not the VI method works equally well for identical potential structures crossing the spacecraft at different angles. To investigate this, we prescribe a potential given by Eq.~(\ref{eq:EH}) with $L_\parallel=100$m, $L_\perp=5L_\parallel=500$m, and perform the calculations for all propagation angles $\theta\in[0,90]^\circ$, $\varphi\in[0,45]^\circ$ with an angular resolution of $\Delta\theta=\Delta\varphi=1^\circ$. For each run, we compute the ratio between the estimated and prescribed speeds $v/v'$ (Fig.~\ref{fig:V_L_stats}a) and the angle between the estimated and prescribed velocities $\alpha=\arccos{(\hat{v}\cdot\hat{v}')}$ (Fig.~\ref{fig:V_L_stats}b). The analysis shows that when $\mathbf{k}$ is oblique to all probe pairs (i.e. $\theta\sim45^\circ$, $\varphi\sim45^\circ$), the velocity estimated by the VI method is very accurate, $\mathbf{v}\approx\mathbf{v}'$. However, as $\mathbf{k}$ becomes close to aligned with one probe pair, and consequently nearly perpendicular to another, the VI method yields large errors with $v\ll v'$, and $\mathbf{v}$ approximately $90^\circ$ from $\mathbf{v}'$. The reason for this is shown in Fig.~\ref{fig:V_L_stats}c, where we plot $V_3$ and $-V_4$ for the case of $\theta=\varphi=10^\circ$, marked by the magenta dot in Figs.~\ref{fig:V_L_stats}a,b. The time delay between $V_3$ and $-V_4$ is much greater than that between the potentials due to the fact that $\Phi(\mathbf{r}_\text{sc})$ is asymmetrically shifted with respect to $\Phi(\mathbf{r}_3)$ and $\Phi(\mathbf{r}_4)$, as shown in the inset where $\Phi(\mathbf{r}_\text{sc})$ is closer to $\Phi(\mathbf{r}_4)$ for $t<0$ and closer to $\Phi(\mathbf{r}_3)$ for $t>0$, causing a large phase shift upon subtraction. For a plane wave, this problem does not appear, indicating that this is a problem related to the 3-dimensional structure of the solitary wave. 

To investigate such 3-dimensional effects, we repeat the statistical analysis for two values of $L_\parallel\in\{50,200\}$ m, and a range of $L_\perp/L_\parallel\in[1,15]$. To quantify how well the method works for the different length scales, we, for each run, calculate the fraction of data where $v/v'$ is larger than some constant $C\in\{0.5,0.7,0.9\}$, and the fraction of data where $\alpha$ is less than some threshold $\alpha_c\in\{10^\circ,20^\circ,30^\circ,40^\circ\}$ (see Fig. S1 for the corresponding $v/v'$ and $\alpha$ surfaces). The results, presented in Fig.~\ref{fig:V_L_stats}d,e show that it is $L_\perp/L_\parallel$, and not the absolute values of $L_\parallel$ or $L_\perp$, that matters. The $L_\perp/L_\parallel$ dependence is not due to the any failure of Eq.~(\ref{eq:V_vel}), but rather it is due to the phase shift observed in the voltages. Using the time delays of $\Phi$ rather than $V$ in Eq.~(\ref{eq:V_vel}) produces accurate results independent of $L_\perp/L_\parallel$. For $L_\perp/L_\parallel<5$ we also see that the quality of the velocity estimate decreases rapidly as $L_\perp/L_\parallel$ decreases. For larger ratios (quasi 1-dimensional structures) however, the VI method generally produces accurate results.

It is important to note that the asymmetry observed in the voltage signals in Fig.~\ref{fig:V_L_stats}c (cf. Fig.~\ref{fig:Vmethod}a) is a strong indicator that they are phase shifted with respect to the potentials. We find that the VI method produces its worst results when the voltages are extremely distorted and poorly correlated. This means that we can identify when the method is unlikely to produce accurate results, using, for example, the ratio of peak amplitudes as a measure of asymmetry. Taking the same potential parameters as for Figs.~\ref{fig:V_L_stats}a,b, but setting $\Delta t_{ij}=0$ when either of $\max(V_i)$ and $\max(-V_j)$ is more than $50\%$ larger than the other, we significantly improve the results as shown in Fig.~\ref{fig:V_L_stats}f,g. These results stress the importance of applying timing analysis only to voltage signals which are very similar to each other.

\begin{figure}
\noindent\includegraphics[width=\textwidth]{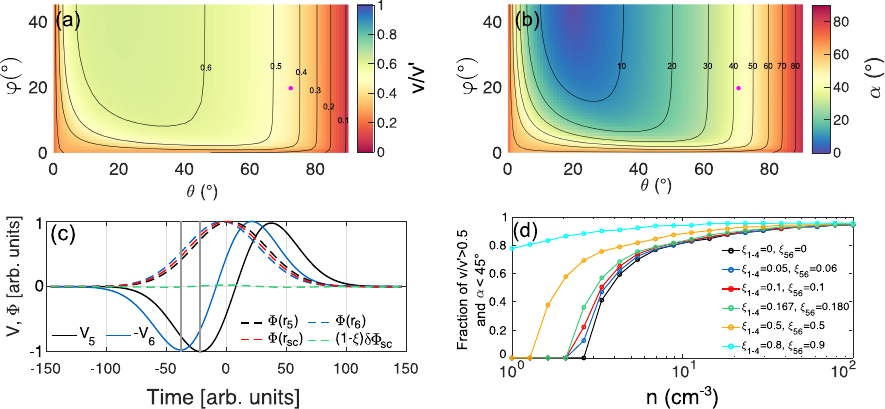}
\caption{Numerical results for voltage interferometry including spacecraft potential effects. (a-b) $v/v'$ and $\alpha$ as a function of $\theta$ and $\varphi$ for $\Phi_0=1$ V, $L_\parallel=100$ m, $L_\perp/L_\parallel=50$, $n=4.3$ cm$^{-3}$, $\xi_{1-4}=0.05$, and $\xi_{5,6}=0.06$. (c) Example of time series of $V_5$ and $V_6$ corresponding to $\theta=70^\circ, \varphi=20^\circ$ (magenta dot in (a,b)). (d) The fraction of points in (a) and (b) which simultaneously fulfill $v/v'>0.5$ and $\alpha<45^\circ$ as a function of density for six combinations of $\xi_{1-4}$ and $\xi_{5,6}$.}
\label{fig:V_xi_stats}
\end{figure}
The above results are representative only of very dense plasma where the spacecraft potential change due to the wave is negligible. In the following, we extend the analysis to more general plasmas by including various values for $n$, $\xi_{1-4}$, and $\xi_{5,6}$. We choose the photoelectron temperature $T_\gamma=5$ eV, and a constant wave amplitude $\Phi_0=1$ V, which can be representative of solitary waves at the magnetopause and bow shock~\cite{graham2016,steinvall2019a,wang2021}. To isolate the spacecraft potential effects, we take a potential that is approximately 1-dimensional, $L_\perp/L_\parallel=50$, with a parallel length scale $L_\parallel=100$ m, consistent with the $L_\parallel$ values reported by \citeA{graham2016} at the magnetopause. Sweeping again over $\theta$ and $\varphi$, we construct similar $v/v'$ and $\alpha$ surface plots as in Fig.~\ref{fig:V_L_stats}. In Figs.~\ref{fig:V_xi_stats}a,b we show an example of these surfaces for the $n=4.3$ cm$^{-3}$, $\xi_{1-4}=0.05$, $\xi_{5,6}=0.06$ case. Clearly, the inclusion of spacecraft potential effects reduces the accuracy of the VI method, causing it to underestimate the speed ($v/v'<0.7$) and rotate $\hat{k}$ by a sometimes significant $\alpha$. Similar surfaces for different values of $n$ and $\xi$ are presented in Figs. S2 and S3 in the supporting information. In contrast to the 3-dimensional effects, where the errors were primarily concentrated near the edges of the surface (Figs~\ref{fig:V_L_stats}a,b), the errors due to spacecraft potential effects appear over the whole surface, i.e. for all propagation directions. In addition, the $V$-signals are not necessarily distorted while producing erroneous velocities. This is exemplified in Fig.~\ref{fig:V_xi_stats}c, where we present the time series of $V_5$ and $V_6$ for the $\theta=70^\circ$, $\varphi=20^\circ$ case of Figs.~\ref{fig:V_xi_stats}a,b. The observed time delay between $V_5$ and $-V_6$ is much greater than that between the potentials, resulting in an erroneous velocity. Importantly, since the voltages are not noticeably distorted in this case (cf. Fig~\ref{fig:V_L_stats}c), we cannot visually identify when the VI method will produce errors due to spacecraft potential effects. In the more extreme cases with larger errors, the voltage signals become distorted. Similar problems can be found for the other probe pairs depending on the various parameters.

To quantify the effects of different combination of $n$ and $\xi$, we sweep over all angles $\theta$ and $\varphi$ in the same way as before, compute the fraction of estimated velocities satisfying both $\alpha\leq45^\circ$ and $v/v'\geq0.5$ for $n\in[1,100]$ with various combinations of $\xi_{1-4}$ and $\xi_{5,6}$. The results, presented in Fig.~\ref{fig:V_xi_stats}d, show that the VI method is very sensitive to both $n$ and $\xi$. If the plasma is dense ($n\gtrsim30$ cm$^{-3}$), the VI method is accurate and practically independent of $\xi$. In more tenuous plasmas ($n\lesssim5$ cm$^{-3}$) however, the VI method is very sensitive to $\xi$, and it produces accurate results only when $\xi$ is large, $\xi\gtrsim0.8$. The fact that the results are accurate when $\xi$ is large follows from Eq.~(\ref{eq:V}), since the $(1-\xi_i)\delta\Phi_\text{sc}$ contribution to $V_i$ becomes negligibly small as $\xi_i\rightarrow1$. Of particular practical interest is the values of $\xi_{1-4}$ and $\xi_{5,6}$ corresponding to the MMS spacecraft. Using the MMS estimate of $\xi_{1-4}=0.167$ reported by \citeA{graham2018b} in the magnetosphere, and assuming $\xi_{5,6}=0.180$, we obtain the green curve in Fig.~\ref{fig:V_xi_stats}d. Five corresponding $v/v'$ and $\alpha$ surfaces for varying values of $n$ are shown in the supporting information Fig.~S3. The results show that the VI method is likely to be inaccurate when applied to MMS data gathered in the magnetosphere, where $n$ tends to be a few particles per cm$^3$ or lower.
Since the $\xi$-parameters are difficult to estimate in-situ, our results highlight the importance of applying the VI method only when the signals are very well correlated, and in a plasma where the spacecraft potential effects can be reasonably be assumed to be small, such as in the dense magnetosheath where $n$ can be up to reach values up to a few $100$ cm$^{-3}$.

\subsection{Reconstructed Electric fields}
\begin{figure}
\noindent\includegraphics[width=0.75\textwidth]{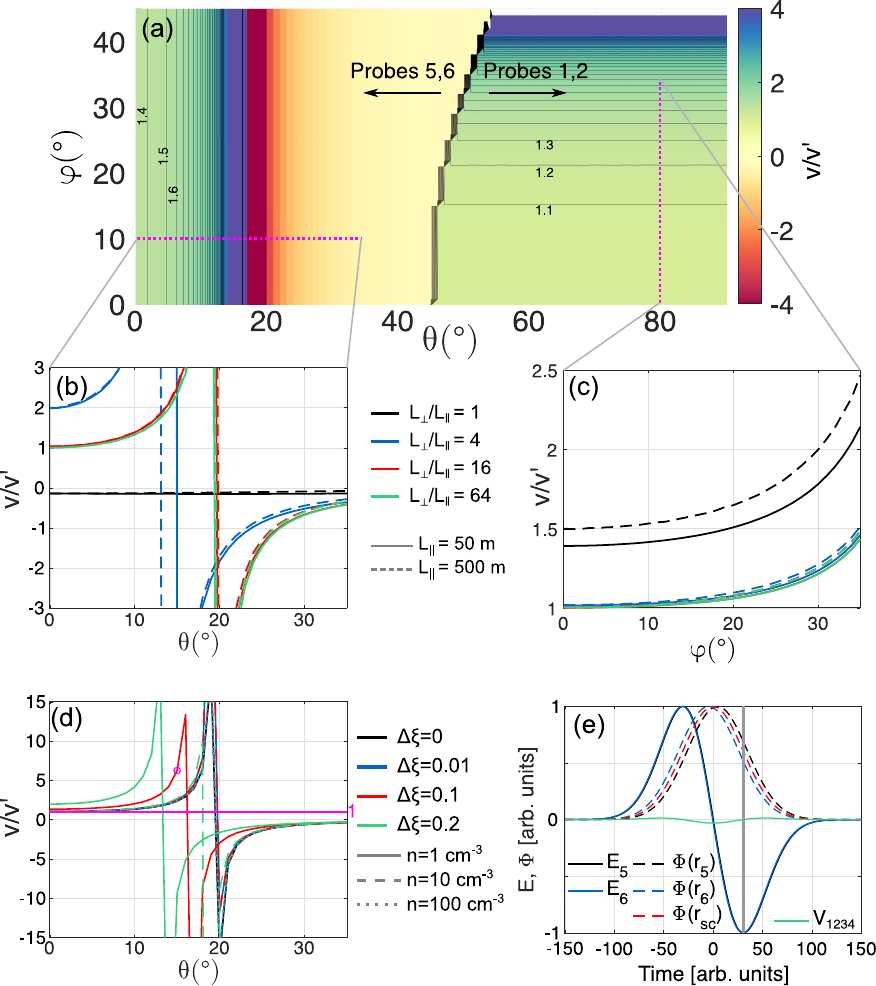}
\caption{Statistics for reconstructed E-field interferometry with $\xi_{1-4}=\xi_{5,6}=0$. (a) $v/v'$ for all angles, with $L_\parallel=500$ m, $L_\perp=5L_\parallel$. Note that the colorbar saturates in the dark purple and red regions, and the contours stop at $v/v'=4$. (b) $v/v'$ as a function of $\theta$ for various $L_\perp/L_\parallel$ with two values for $L_\parallel$. The data is taken at $\varphi=10^\circ$, marked by the horizontal magenta line in (a). (c) Same format as (b), but with data from the vertical magenta line in (a); the green and red curves are almost perfectly overlapping. (d) $v/v'$ as a function of $\theta$ for different values of $n$ and $\Delta\xi$, for $L_\perp/L_\parallel=64$ and $L_\parallel=100$ m. The magenta line shows $v/v'=1$. (e) Waveform of $E_5$ and $E_6$ for the $n=1$ cm$^{-3}$, $\Delta\xi=0.1$, $\theta=15^\circ$ case in (d), marked by the magenta circle.}
\label{fig:E60stats}
\end{figure}

Next, we explore the REI method in a similar way as above, starting with $\delta\Phi_\text{sc}$=0. In Fig.~\ref{fig:E60stats}a we plot $v/v'$, where $v$ is the speed obtained by Eq.~(\ref{eq:v_E60}), as a function of propagation angles for a potential structure with $\Phi_0=1$ V, $L_\parallel=500$ m, $L_\perp=5L_\parallel$. (We will later see that the choice of $L_\parallel$ does not affect the results much.) The surface is split into two regions. The right region corresponds to when $\hat{k}$ is most aligned with probes 1 and 2 ($\theta\gtrsim45^\circ$), and $E_1$ and $E_2$ are used for timing. The left region corresponds to when $\hat{k}$ is most aligned with probes 5 and 6 ($\theta\lesssim45^\circ$), and we use $E_5$ and $E_6$ for timing. 

When the spin-plane probes are used for timing ($\theta\gtrsim45^\circ$), we find that the REI method tends to overestimate the speed ($v\geq v'$). The error starts increasing very slowly with increasing $\varphi$; for example, $v/v'=1.1$ at $\varphi\approx18^\circ$, and $v/v'=1.5$ at $\varphi\approx31^\circ$, before it quickly diverges for larger $\varphi$ (note that the colorbar is limited to $v/v'=\pm4$ to make the relevant details visible). This diverging behavior is caused by phase shifts in $E_1$ and $E_2$ due to $(V_3+V_4)/2$ being a poor estimate of $V_\text{sc}$ when $\hat{k}$ has a significant component along probes 3 and 4 (recall Fig.~\ref{fig:Emethod}b). In contrast to the strong $\varphi$ dependence, we find that $v/v'$ is independent of $\theta$ in this region. This independence is due to two reasons. First, the $\cos{\gamma}$ correction applied in Eq.~(\ref{eq:v_E60}), which accounts for the fact that $\hat{k}$ is not aligned with the probe pair used for interferometry. Second, the fact that only the spin-plane probes are used, which means that they are all affected in the same way by $\theta$ and no phase shift is introduced between $E_1$ and $E_2$. The $\theta$ independence means that it can be preferable to use the spin-plane probes for interferometry instead of the axial probes (discussed next) for values of $\theta<45^\circ$, assuming the time-delay is large enough to be accurately resolved.

When the axial probes are used ($\theta\lesssim45^\circ$), we instead find that $v/v'$ is independent of $\varphi$, but strongly dependent of $\theta$. The independence of $\varphi$ is due to two reasons: The first reason is that the angle between $\hat{k}$ and the axial probe pair, by definition, is unaffected by $\varphi$. The second reason is that we use all four spin-plane probes to estimate $V_\text{sc}$. If only two spin-plane probes are used, a dependence of $\varphi$ is obtained due to the reduced symmetry, and consequently a worse $V_\text{sc}$ estimate. The observed $\theta$ dependence of $v/v'$ is expected since the spin-plane estimate of $V_\text{sc}$ becomes inaccurate and phase shifts appear when $\mathbf{k}$ approaches the spin-plane.
The axial REI method is thus most accurate for $\theta=0$ ($\hat{k}$ is aligned with the axial probe), where $v/v'\approx1.4$. We find that $v/v'$ grows to infinity (corresponding to $\Delta t_{56}=0$) at $\theta\approx17^\circ$. For $\theta>17^\circ$, the structure appears to move in the opposite direction. We can understand this apparent propagation direction reversal by imagining that a potential structure is moving towards the spacecraft in such a way that it will pass probe 5 before probe 6. We would then find $\Delta t_{56}=0$ (corresponding to an infinite speed) if $E_5=E_6$. In terms of the voltages, this corresponds to $V_\text{sc}-V_5=V_6-V_\text{sc}$, or $V_5+V_6=2(V_1+V_2+V_3+V_4)/4$. Since the axial and spin-plane booms are of different length, this expression can be satisfied at different angles for different $\Phi$ profiles. For sinusoidal plane waves, this reversal occurs at $\theta\approx20^\circ$. The key point here is that $V_{1234}$ can be a poor reference value when computing $E_5$ and $E_6$ for certain $\theta$. While there are clear limitations to this method, it tends to work well when the wavevector is close to aligned with the probe pair (i.e. small $\gamma$).

In order to investigate the effects of $L_\parallel$ and $L_\perp$, we define two cuts on the surface in Fig.~\ref{fig:E60stats}a, $\varphi=10^\circ$, and $\theta=80^\circ$ (magenta dotted lines), and plot $v/v'$ from these cuts as a function of $\theta$ (Fig.~\ref{fig:E60stats}b) and $\varphi$ (Fig.~\ref{fig:E60stats}c) for different values of $L_\parallel$ and $L_\perp/L_\parallel$. For the axial probes (Fig.~\ref{fig:E60stats}b), we find a large variability for small values of $L_\perp/L_\parallel$. Importantly, when $L_\perp/L_\parallel=1$, the axial REI method never produces the correct sign of $\hat{k}$. As we go to large $L_\perp/L_\parallel$, the method is accurate up to $\theta\approx10^\circ$, after which $v/v'$ rapidly diverges. The angle at which $v\rightarrow\infty$ (or equivalently $\Delta t_{56}\rightarrow0$) converges to $\approx20^\circ$ as $L_\perp/L_\parallel$ increases. The results from the spin-plane probes (Fig.~\ref{fig:E60stats}c) are less sensitive to $L_\perp/L_\parallel$, and quickly approach the asymptotic curve at $L_\perp/L_\parallel\approx4$. In summary, the REI method yields accurate results when $\theta\lesssim10^\circ$ if the axial probes are used and the potential structure is approximately 1-dimensional, and when $\varphi\lesssim30^\circ$ if the spin-plane probes are used.

Since probes 1-4 are identical, this method is completely independent of $\delta\Phi_\text{sc}$ when only these probes are used. The results in Fig.~\ref{fig:E60stats}c and the right-hand part of Fig.~\ref{fig:E60stats}a thus also hold when $\delta\Phi_\text{sc}\neq0$. However, when $E_5$ and $E_6$ are used for timing, this is no longer the case since the spin-plane probes are used to compute $V_\text{sc}$. The consequences of spacecraft potential effects are shown in Fig.~\ref{fig:E60stats}d, where we plot $v/v'$ along the horizontal cut for different values of $n_e$ and $\Delta\xi=\xi_{5,6}-\xi_{1-4}$, using $L_\parallel=100$ m and $L_\perp/L_\parallel=64$. The results reveal a strong dependence on $\Delta\xi$, when the density is of the order of $n\sim1$ cm$^{-3}$. For denser plasmas ($n>10$ cm$^{-3}$), the $\Delta\xi$ dependence is practically gone. It is not possible to determine whether or not the analysis will be accurate by looking at the $E_5$ and $E_6$ signals, as they remain undistorted. This is illustrated in Fig.~\ref{fig:E60stats}e, where we plot $E_5$ and $E_6$ together with $\Phi(\mathbf{r}_5)$ and $\Phi(\mathbf{r}_6)$ for the $n=1$, $\Delta\xi=0.1$, $\theta=15^\circ$ case, marked by the magenta circle in Fig.~\ref{fig:E60stats}d. In this case, $E_5$ and $E_6$ are almost perfectly overlapping, resulting in an overestimated velocity. The fact that the axial and spin-plane booms are different for real spacecraft can also affect the axial REI method. Different boom types have different frequency responses, which means that the $E_5$ and $E_6$ signals can be artificially distorted (since spin-plane and axial probe data are mixed), and the observed $\Delta t_{56}$ can potentially be affected. This boom-response effect is not included in our model, but it most likely has a small effect compared to the other errors discussed above. 
From our analysis we conclude that the axial REI method is expected to produce accurate results for $\theta\lesssim10^\circ$ in dense ($n\gtrsim10$ cm$^{-3}$) plasmas. For more tenuous plasmas, the accuracy of the axial REI method strongly depends on $\Delta\xi$, it should only be used if there are reasons to assume that $\Delta\xi$ is small ($\Delta\xi\lesssim0.1$). 

\subsection{Diagonal E-fields}
\begin{figure}
\noindent\includegraphics[width=\textwidth]{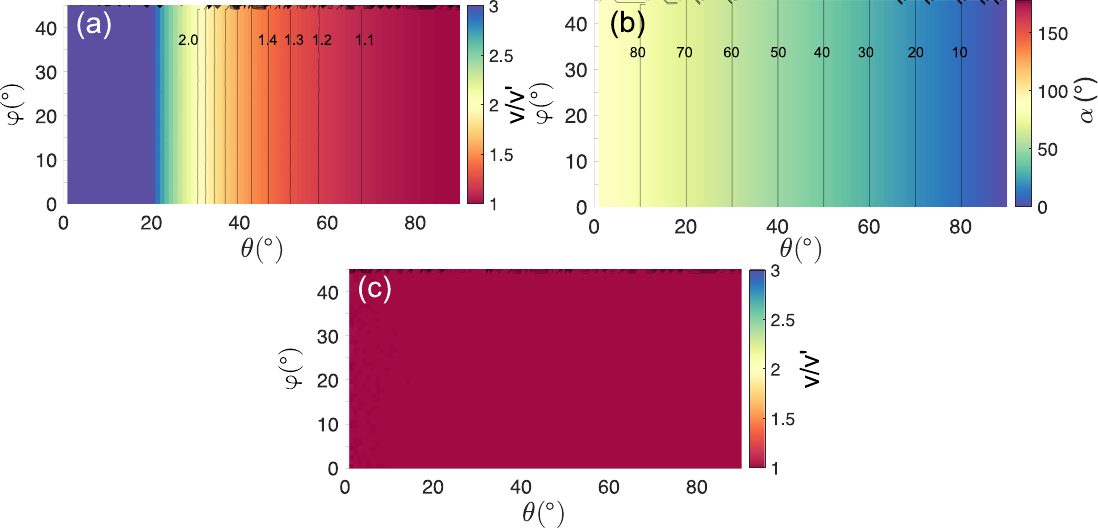}
\caption{Statistical results for the diagonal E-field method. (a) Ratio of $v/v'$ where $v$ is given by Eq.~(\ref{eq:v_E85}). Note that the color map is saturated for $\theta\lesssim20^\circ$, corresponding to $v/v'\geq3$. (b) Angle $\alpha$ between $\mathbf{v}$ and $\mathbf{v}'$. (c) Same as (a), but with $\hat{k}$ known, and $v\rightarrow v\sin\theta$.}
\label{fig:Ediagstats}
\end{figure}

Finally, we explore the DEI method. We repeat the same sweeps over $\theta$ and $\varphi$ as before, with $L_\parallel = 100$ m, $L_\perp=5L_\parallel$ (the choice of $L_\parallel$ and $L_\perp$ does not affect the results). As shown in Fig.~\ref{fig:Ediagstats}a,b, we find that this method produces accurate results when $\hat{k}$ is in the spin-plane (i.e. $\theta=90^\circ$).
As $\hat{k}$ becomes more oblique, the estimated velocity $\mathbf{v}$ starts to differ from the prescribed $\mathbf{v}'$, and we find that the angle between the two velocities is given by $\alpha=90^\circ-\theta$. This is due to the $\hat{z}$ component of $\hat{k}$ not being accounted for in Eq.~(\ref{eq:v_E85}). If $\hat{k}$ is known, we can account for this by multiplying $v$ by $\sin{\theta}$ (Fig.~\ref{fig:Ediagstats}c), resulting in perfect agreement for almost all angles, the two exceptions being $\varphi=45^\circ$, and $\theta=0^\circ$. When $\varphi=45^\circ$, $E_{13}=E_{42}=0$, and $\hat{k}$ is perpendicular to the separation vector of $E_{41}$ and $E_{23}$, causing the method to break down as mentioned in Sec.~\ref{sec:DEI_intro}. The same reasoning applies for the $\theta=0^\circ$ case, where the motion is purely along the axial direction, and no time delay is observed in the spin-plane. In practice, due to noise and limited temporal resolution, this will also be a problem for angles close to these special cases. These results are independent of $L_\parallel$ and $L_\perp$, and the method works equally well for sinusoidal plane waves. Since the DEI method only uses the identical spin-plane probes, it is unaffected by spacecraft potential effects, making this a suitable method for all plasma conditions.

\section{Implications for Analysis of in-situ Measurements}
\label{sec:spacecraft}
In the previous section we showed that the VI method is useful when the potential structure is approximately 1-dimensional, and spacecraft potential effects are negligible. When this is not the case, the voltage signals tend to be distorted, and the results are inaccurate. Spacecraft potential effects can be avoided by using electric field based methods instead. Our results show that the DEI method is always preferable to the REI method, as it always works in the spin-plane except at $\varphi=45^\circ$, where the REI method is highly inaccurate. If $\hat{k}$ is known (a necessary requirement for the REI method), the DEI method gives accurate speeds for all propagation directions, whereas REI is only accurate when $\hat{k}$ is close to aligned with one of the probe pairs. Next, we apply the VI and DEI methods on in-situ spacecraft data to illustrate when they are applicable. We use data from the electric field double probe instrument \cite{lindqvist2016,ergun2016} on the MMS spacecraft. The instrument is in the high cadence \emph{burst mode}, where each probe voltage is sampled at a rate of 8192 samples per second. It is important to note that the different probe potentials are sampled sequentially, not simultaneously, which was the case in our model. It can be important to correct for this if the observed time-delays are very small (of the order of $\sim 10$ $\mu$s). One method for doing this correction is described in Appendix D of the EDP Data Products User Guide~\cite{edp_dataproduct}. For the examples we present below, this time correction has a negligible effect on the estimated velocity vectors. In the following, we present two examples of solitary waves observed by the MMS spacecraft in different plasma regions, and we discuss a few complications and limitations that appear when we start using real data.

\begin{figure}
\noindent\includegraphics[width=\textwidth]{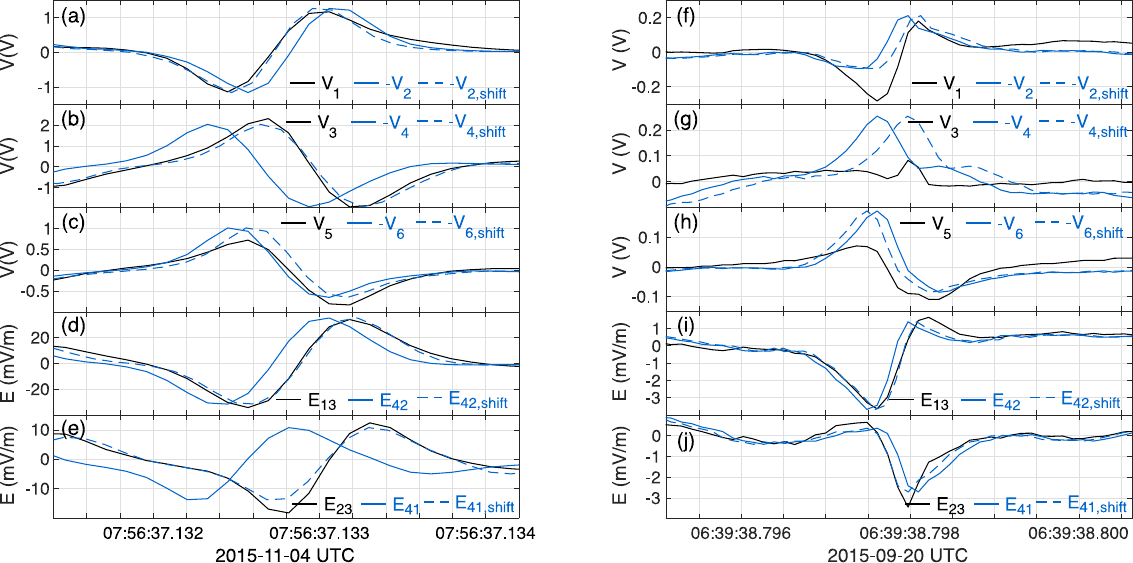}
\caption{Example observation of two solitary waves observed by MMS1 (left column) and MMS3 (right column). (a-c) Voltage signals. (d,e) Diagonal electric field signals. The dashed curves have been time-shifted. (f-j) Same format as (a-e).}
\label{fig:mms_example}
\end{figure}

In the first example, presented in Fig.~\ref{fig:mms_example}a-e, MMS1 observed an ion hole in the downstream side of the Earth's bow shock, where the electron density and temperature were $n=35$ cm$^{-3}$, and $T_e=94$ eV, respectively, corresponding to a Debye length of $12$ m. One can see that the voltage signals for the spin-plane probes are very well correlated (Fig.~\ref{fig:mms_example}a,b), whereas $V_5$ and $-V_6$ are slightly asymmetric, similar to the voltages in Fig.~\ref{fig:V_L_stats}c, suggesting that the axial timing might give inaccurate results.
The time delays, obtained by cross-correlation analysis are $\Delta t_{12}=-0.10$ ms, $\Delta t_{34}=0.32$ ms, and $\Delta t_{56}=0.11$ ms.
The corresponding spin-plane and complete phase velocity vectors are $\mathbf{v}^{(V)}_\text{spin}=180[-0.30,0.95,0]$ and $\mathbf{v}^{(V)}=110[-0.18,0.58,0.80]$ km/s, respectively. The spin-plane phase velocity is in good agreement with that obtained by the diagonal E-field interferometry (using $\Delta t_{42,13}=0.17$ ms, $\Delta t_{41,23}=0.48$ ms) $\mathbf{v}^{(E)}_{\text{spin}}=170[-0.35,0.94,0]$, differing only by $8^\circ$ in direction, and $6\%$ in magnitude. This agreement supports the conclusion from the numerical study that the VI method produces accurate results in dense plasmas when the voltage signals are well correlated and undistorted. 
The ambient magnetic field direction is $\hat{B}=[-0.28,0.66,0.70]$ (corresponding to $\theta=45^\circ$, $\varphi=113^\circ$), and the angle between $\hat{B}_\text{spin}$ the two $\mathbf{v}_\text{spin}$ estimates are $6^\circ$ and $2^\circ$, for $\mathbf{v}_\text{spin}^{(V)}$ and $\mathbf{v}_\text{spin}^{(E)}$ respectively. Such small angles are consistent with field-aligned propagation in the spin-plane. Moreover, the full $\hat{v}^{(V)}$ vector is $9^\circ$ from $\hat{B}$. It is possible that the slightly increased deviation from field-aligned propagation when including the axial direction, is due to a phase shift between $V_5$ and $-V_6$, as hinted by the asymmetry in the signals (Fig.~\ref{fig:mms_example}c). If we assume $\hat{k}\parallel\hat{B}$, we find the phase-speed according to the diagonal E-field method to be $v^{(E)}_\text{spin}\sin\theta=120$ km/s, comparable to $v^{(V)}=110$ km/s. This example shows that well correlated voltage signals enable the voltage interferometry method to produce accurate results, but also that voltage asymmetries may introduce errors. Moreover, under the assumption of field-aligned propagation (which is not always valid for ion holes \cite{wang2021}), the two methods produce very similar results, supporting the validity of the assumption and the accuracy of $\mathbf{v}^{(V)}$.

The second example (Fig.~\ref{fig:mms_example}f-j) is an electron hole observed by MMS3 during a magnetopause crossing where $n=11$ cm$^{-3}$ and $T_e=67$ eV, corresponding to a Debye-length of 18 meters. Previous multi-spacecraft analysis has shown that such electron holes exhibit field-aligned propagation~\cite{steinvall2019a}. In this case, the voltage signals (f-h) are significantly distorted and poorly correlated, while the diagonal E-fields are well correlated. Naively applying the VI method (with $\Delta t_{12}=0.15$, $\Delta t_{34}=0.36$, $\Delta t_{56}=-0.12$ ms) in the spin-plane gives $\mathbf{v}^{(V)}_\text{spin}=150[0.38,0.92,0]$ km/s, which is $89^\circ$ from the projection of the magnetic field $\hat{B}\approx[-0.22,0.09,0.97]$ (corresponding to $\theta=14^\circ$, $\varphi=156^\circ$) onto the spin plane. Including the third dimension gives $\mathbf{v}^{(V)}=97[0.24,0.58,-0.78]$, which is $83^\circ$ from $-\hat{B}$. The DEI method on the other hand gives (using $\Delta t_{42,13}=0.10$, $\Delta t_{41,23}=-0.12$ ms) $\mathbf{v}^{(E)}_{\text{spin}}\approx540[0.99,-0.09,0]$ km/s, only $18^\circ$ from $-\hat{B}_\text{spin}$, close to the expected field-aligned propagation. 
Multiplying $\mathbf{v}^{(E)}_{\text{spin}}$ by $\sin\theta$ to account for the axial velocity component, we find ${v}^{(E)}=130$ km/s.
From the DEI results we see that the spin-plane projection of $\hat{k}$ is close to perpendicular to probes 3 and 4. This can partly explain why the VI method produces such erroneous velocities. Recall in Fig.~\ref{fig:V_L_stats}a,b, that when $\varphi$ is very small (as in this case), VI is expected to give an angular error of $\alpha\approx90^\circ$ and to underestimate the speed of non-planar structures. If we, as in Fig.~\ref{fig:V_L_stats}f,g, set $\Delta t_{34}=0$, we obtain a better estimate $\mathbf{v}^{(V)}=120[0.30,0,-0.95]$ using VI. However, as the other voltages are also poorly correlated, it is possible that spacecraft potential effects are also important.

While we have incorporated the effect density perturbations have on the spacecraft potential in our model, we have neglected the fact that the E-field of the waves can affect the spacecraft potential by enhancing the photoelectron escape~\cite{torkar2017,graham2018b}. This effect in principle also applies to the small photoelectron cloud around the probes, but due to the probe biasing and the much larger spacecraft photoelectron cloud, it primarily modulates the spacecraft potential fluctuations $\delta\Phi_\text{sc}$. For this reason, such enhanced photoelectron escape will primarily affect the voltage interferometry, assuming such effects are symmetric with respect to the spacecraft body. If an asymmetry is present, it can also affect the electric field, possibly influencing the E-field interferometry methods. However, for the above examples this appears to have had a negligible effect on the E-fields since they are well correlated.

One additional limitation of our results is the fact that we have assumed that the wavelength is large compared to the probe-to-spacecraft separation. This might pose some limitations on wave analysis in plasmas with a very short Debye length, where waves with wavelengths shorter than the probe separation can exist. The main effect short wavelengths have on these methods is to distort the waveform (due to spatial aliasing). However, this issue is not in principle due to the interferometry methods, but rather the spacecraft geometry, so the three methods are affected essentially in the same way. For sinusoidal waves of short wavelength interferometry is problematic because the timing becomes ambiguous as one does not know which wave period is to be timed against which. But for solitary waves it is not the length scale $L_\parallel$ that matters, but rather the spacing between subsequent solitary waves. If the spacing is large enough so that each solitary wave can completely pass the spacecraft before the next one arrives, then no matter how short $L_\parallel$ is (given that it is large enough to be detected), it will unambiguously appear as two peaks in the data. However, if the spacing is sufficiently short so that several solitary waves fit between the probe and spacecraft, an ambiguity appears. A secondary effect short length scales have is that the measured E-field components are attenuated~\cite{goodrich2018}. The difference in boom length between the axial and spin-plane probes means that the spin-plane E-fields is more strongly reduced in amplitude on one of the probes, which shifts the apparent direction of the field. This is a problem when the direction of $\mathbf{E}$ is important, such as when maximum variance is used to determine the propagation direction needed for the REI and DEI methods. For these two reasons, one needs to be careful when applying single-spacecraft interferometry on waves which likely have short length scales.

We mentioned in Sec.~\ref{sec:data_theory} that the antenna shortening effect~\cite{pedersen1998} only affects the amplitude of $E_{ij}$, and not the time-delay obtained from interferometry. While this is true, the shortened antenna length affects the velocity estimate by changing the spatial baseline. However, this affects all methods in the same way, and is therefore not of interest in our comparative analysis. Moreover, based on experience from the Cluster mission which is equipped with a similar electric field instrument in the spin-plane~\cite{gustafsson1997}, we expect $L_{\text{eff}}$ to differ from the geometric antenna length by only a few tens $\%$ for MMS~\cite{khotyaintsev2014}. This effect is therefore likely much less important than the errors inherent in the methods. 

It is possible to reduce the effect density perturbations have on the spacecraft potential by utilizing active spacecraft potential control (ASPOC),~\cite<e.g.>[]{torkar2016}. The principle of ASPOC is to emit an ion beam from the spacecraft, corresponding to a positive current leaving the spacecraft. This has the effect of reducing the relative importance of $I_e$ in determining $\Phi_\text{sc}$, which means that $\Phi_\text{sc}$ is less affected by the density perturbations associated with waves. For this reason, using ASPOC could improve the performance of the VI method.
For the case of MMS, the effect ASPOC has on the current balance is negligible when the spacecraft is in the dense magnetosheath and bow shock, where the VI method is already accurate. The ASPOC current is more important in the tenuous magnetotail, and the VI method is therefore likely improved if ASPOC is turned on. However, since the length scales of waves in the magnetotail tend to be large ($\sim$ km) compared to the probe separation, it is very difficult to measure a time delay between individual probes. Under such conditions the multi-spacecraft interferometry is a preferable method to use.

Finally, since our analytical model contains several simplifying assumptions and idealizations, we stress that our results show how well the interferometry methods work in the ideal case. Our results thus provide an upper limit on the accuracy of the methods. Since other complications are affecting the in-situ data (e.g. a finite instrument sampling time, rapid changes in the photoelectron cloud, etc.), the methods are expected to perform worse than presented here. However, as illustrated by the example observations of MMS discussed above, these additional complications are not always important, and the interferometry methods can, under the right circumstances, be applied to in-situ data with high accuracy.

Our results have been derived for the case of MMS, but the results are likely also applicable in a qualitative sense to other similar spacecraft such as Cluster and THEMIS. For spacecraft that are significantly different such as Parker Solar Probe or Solar Orbiter, there may be larger differences, but nevertheless the main qualitative conclusions, in particular regarding the complications of the voltage interferometry, should still apply.

\section{Conclusions}
\label{sec:summary}
In summary, we have used an analytical model to analyze the accuracy of three single-spacecraft interferometry methods. By creating synthetic time series data corresponding to data from the electric field probes on the MMS spacecraft, we compare the prescribed velocity of an electrostatic wave crossing the synthetic spacecraft with the one obtained from the interferometry methods. In particular, we investigate how the accuracy of the interferometry methods is affected by the shape, specifically the three-dimensionality, of the electrostatic potential structure, the direction with which it crosses the spacecraft, and the density perturbations associated with it. We summarize our findings for the different methods below.

The voltage interferometry (VI) method is accurate when spacecraft potential fluctuations due to the observed wave are negligible. For the probe geometry of MMS, and when ASPOC is not operating, we find that this is the case in very dense plasmas ($n\gtrsim10$ cm$^{-3}$) depending on wave properties. In such cases, the method works best for 1-dimensional potential structures, or, for three-dimensional structures when the wavevector $\mathbf{k}$ is not close to perpendicular to any probe pair (Figs.~\ref{fig:V_L_stats}a,b,d,e). If $\mathbf{k}$ is close to perpendicular to a probe pair, the voltages will be distorted in an identifiable way (Fig.~\ref{fig:V_L_stats}c). By disregarding the time delay between heavily distorted signals, i.e. setting the corresponding $\Delta t_{ij}=0$, the results can be significantly improved (Figs.~\ref{fig:V_L_stats}f,g). When spacecraft potential effects are important, typically in low-density $n\lesssim10$ cm$^{-3}$ plasmas, the method quickly starts producing erroneous velocities (Fig.~\ref{fig:V_xi_stats}) even without visible distortion of the signals. Finally, we stress that the VI method should only be applied to voltage signals that are close to identical, since distorted voltage signals are indicative that VI interferometry will yield significant velocity errors in both speed and direction.

The reconstructed E-field method is accurate in the spin-plane when the direction of $\mathbf{k}$, $\hat{k}$, is known, and is close to one the probe pairs ($\varphi\lesssim30^\circ$, see Fig.~\ref{fig:E60stats}a,c). It is also independent of spacecraft potential effects, making it suitable in most magnetized plasma conditions. When used along the axial direction, the method requires $\theta$ to be very small ($\lesssim10^\circ$; Fig.~\ref{fig:E60stats}a,b), and it is sensitive to spacecraft potential effects, meaning it should only be used in plasmas where such effects can safely be assumed to either be negligible, or affect the axial and spin-plane probes in the same way, i.e. $\xi_{1-4}=\xi_{5,6}$, where $\xi_i$ is the fraction of the spacecraft potential observed by probe $i$. 

The diagonal E-field method is accurate in the spin-plane, except near $\varphi=45^\circ$, where it breaks down. If $\hat{k}$ is known, projecting the spin-plane phase velocity onto $\hat{k}$ yields an accurate velocity.

We conclude that the diagonal E-field interferometry method is almost always preferable to the other two methods for determining the spin-plane component of the phase velocity. Because of this, this method should be used as a reference to see if the voltage interferometry method is accurate in the spin-plane. If $\hat{k}$ is unknown, voltage interferometry shall be used to find $v_z$ only if the signals are very well correlated and the plasma is dense. The reconstructed E-field method shall be used to find $v_z$ only when $\hat{k}$ is within $\lesssim20^\circ$ from the spin-axis, and there is reason to assume that the spacecraft potential effects are negligible. If $\hat{k}$ is known, the diagonal E-field interferometry method shall be used to find the spin-plane phase velocity and then project it back to $\hat{k}$ to obtain the full velocity vector.

\acknowledgments
We thank I. Y. Vasko, R. Wang and F. Mozer for fruitful discussions at an early stage of this study.
This work is supported by the Swedish National Space Agency, Grant 128/17, and the Swedish Research Council, Grant 2016-05507. MMS data are available at \\ \noindent{https://lasp.colorado.edu/mms/sdc/public/.}

%
%

\bibliography{biblio.bib}

%
%
%
%
%

\end{document}